\documentclass[aps,prb,preprint,nofootinbib]{revtex4-1}
\usepackage{amsmath}  
\usepackage{amsfonts} 
\usepackage{graphicx} 
\usepackage{color}
\usepackage[pdftex]{hyperref}

\newcommand{\di}{\mathrm{d}}
\def\pa{\partial}

\makeatletter 
\newcommand\footnoteref[1]{\protected@xdef\@thefnmark{\ref{#1}}\@footnotemark}
\makeatother

\begin{document}
\title{Counterintuitive effect of gravity on the heat capacity of a metal sphere: re-examination of a well-known problem}

\author{Giacomo De Palma\footnote{The two authors contributed equally to this work. \label{note1}}}
\email{giacomo.depalma@sns.it}
\affiliation{NEST, Scuola Normale Superiore and Istituto Nanoscienze-CNR, I-56127 Pisa,
Italy}
\affiliation{INFN, Pisa, Italy}
\author{Mattia C. Sormani\footnoteref{note1}}
\email{mattia.sormani@physics.ox.ac.uk}
\affiliation{Rudolf Peierls Centre for Theoretical Physics, 1 Keble Road, Oxford}

\begin{abstract}
A well-known high-school problem asking the final temperature of two spheres that are given the same amount of heat, one lying on a table and the other hanging from a thread, is re-examined. The conventional solution states that the sphere on the table ends up colder, since thermal expansion raises its center of mass. It is found that this solution violates the second law of thermodynamics and is therefore incorrect. Two different new solutions are proposed. The first uses statistical mechanics, while the second is based on purely classical thermodynamical arguments. It is found that gravity produces a counterintuitive effect on the heat capacity, and the new answer to the problem goes in the opposite direction of what traditionally thought.
\end{abstract}

\maketitle

\section{Introduction and statement of the problem}
In the very first International Physics Olympiad in 1967, the following problem\cite{IPhOsite,IPhO1,IPhO2} was assigned:

{\it ``Consider two identical homogeneous balls, $A$ and $B$, with the same initial temperatures. One of them is at rest on a horizontal plane, while the second one hangs on a thread. The same quantities of heat have been supplied to both balls. Are the final temperatures of the balls the same or not? Justify your answer. (All kinds of heat losses are negligible.)''}

\begin{figure}[h]
\centering
\includegraphics[width=0.5\textwidth]{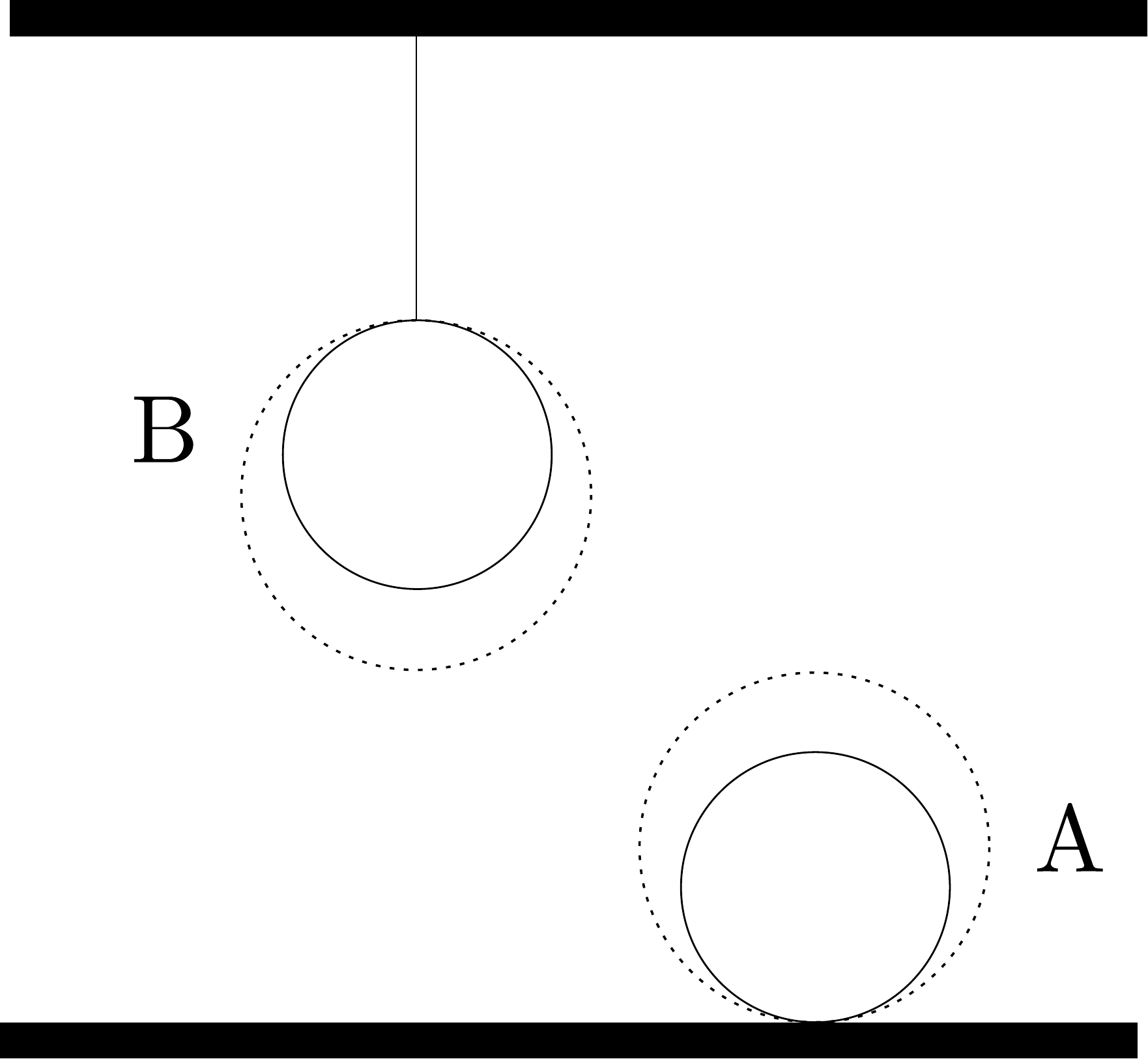}
\caption{The setup of the problem.}
\label{fig}
\end{figure}

See Fig. \ref{fig}. This problem is now well-known and can be found in several problem books and collections.\cite{200puzzling,madaboutphysics,AIEEE,dukechallenges,caltech,princeton} The main reason for its popularity is that the solution usually presented, while being tricky to find, is very simple and elegant, and uses tools accessible to a high-school student. In this paper we show that this solution, which we will refer to as the conventional solution, is incorrect as it violates the second law of thermodynamics. We propose a new result and we derive it using two rather different approaches: first through statistical physics, and second through only the good old nineteenth-century classical thermodynamics.

\section{The conventional solution}
The most common reaction when one first tries to solve the problem is to be puzzled. Why should there be any difference at all between the balls? All kinds of heat losses, such as thermal conduction with the air and the ground are to be neglected! Closer scrutiny reveals that there are some relevant differences. The conventional solution relies on a difference related to thermal expansion to argue as follows. When heat is supplied, the ball $A$ at rest on the horizontal plane expands, and its center of mass rises. Therefore, the ball $A$ uses part of the heat supplied to increase the gravitational potential energy of the center of mass and will have the lower final temperature. On the contrary, the center of mass of the ball $B$ is lowered by thermal expansion and energy is gained, therefore the final temperature of ball $B$ will be higher.

Let us write down some formulas corresponding to this conventional solution. We are interested in the change in the heat capacity when a constant gravitational field is applied, with respect to the zero-gravity case. Let $C_0$ be the heat capacity of a ball in the absence of gravity. According to the conventional solution, when the ball $A$ is heated, the center of mass raises by an amount $\di R = \alpha R \di T$, where $\di T$ is the increase in temperature, $\alpha$ is the coefficient of thermal expansion and $R$ is the radius of the ball. The ball gains a potential energy $\di \Phi = m g \di R$, where $m$ is the mass of the ball and $g$ the gravitational acceleration. Thus, when an amount of heat $\delta Q$ is supplied to the system we have:
\begin{equation}
\delta Q = C_0 \di T + M g \di R = C_0 \di T + M g \alpha R \di T = \left(C_0 + M g \alpha R\right) \di T\;.
\end{equation}
We can rephrase this by saying that the heat capacity (in presence of gravity) of the ball $A$ is
\begin{equation}
C_A = C_0 + M g \alpha R\;.
\end{equation}
Following a similar reasoning, the heat capacity for the ball $B$ would be
\begin{equation}
C_B = C_0 - M g \alpha R\;.
\end{equation}
These results amount to saying that the heat capacity $C$ is changed in the presence of a small gravitational field by an amount
\begin{equation}
\frac{\pa C(g,T)}{\pa g}  = M \alpha R\;. \label{conventionalresult}
\end{equation}
For most materials $\alpha>0$, which implies $C_A > C_B$. Therefore the conventional solution says that the final temperature of the ball $A$ is lower than the final temperature of the ball $B$.

\section{The Conventional solution violates the 2nd law of thermodynamics}
What is wrong with the above solution? A first insight can be gained by the following considerations. Assume, as is implicitly done in the conventional solution, that the internal energy $U$ and the radius $R$ of the ball depend only on the temperature $T$ and not on $g$. Consider the following cycle: start with the ball on the the table at some cold temperature $T_1$. Now connect the ball to a warm heat bath at temperature $T_2 = T_1 + \di T >T_1$. According to the conventional solution, the warm bath will transfer an amount of heat $\delta Q_{\rm abs} = \left(C_0 + M g \alpha R\right)\di T$ to the ball, and its center of mass will raise by $\di R = \alpha R \di T$. Now hang the ball from a thread and then remove the table. Since we assumed that $R$ depends only on $T$, the position of the center of mass will not change. Now connect the ball to a cold bath at temperature $T_1$. The ball radius will go back to its original value, and the center of mass will raise again (because this time it is hanging from the thread). In doing this, an amount of heat $\delta Q_{\rm rel} = \left(C_0 - M g \alpha R\right)\di T$ will be transferred from the ball to the cold bath (to first order in $\di T$). Let us calculate the net result of this process. The ball has raised by a total amount $2 \di R =  2\alpha R \di T$. The corresponding gravitational potential energy gain is potentially convertible into work by letting the ball fall. The efficiency of this process is given by the ratio of the work gained divided by the amount of heat absorbed, and therefore is given by
\begin{equation}
\eta = \frac{2 M g \alpha R }{C_0 + M g \alpha R}\;.
\end{equation}
Then $\eta$ does not depend on $\di T$, and in particular it does not tend to zero as $\di T$ tends to zero! A possible way of expressing the second law of thermodynamics is the statement that any cycle working only between the two temperatures $T_1$ and $T_2$ cannot be more efficient than a Carnot cycle working between the same temperatures, whose efficiency is\cite{feynman}
\begin{equation}
\eta_{\rm Carnot} = \frac{\di T}{T_2}\;.
\end{equation}
Therefore, if $\di T$ is small enough, the efficiency of the cycle with the balls becomes bigger than the efficiency of a Carnot cycle. In other words, the conventional solution presented above violates the second law of thermodynamics.

The reader may now ask what happens when one actually tries to perform the cycle described above. We know that the second law cannot be violated in the real world. Up to now, we have always assumed that the balls remain exactly spherical, and that gravity does not affect their radius or their shape. In reality, when the sphere lies on the table at thermal equilibrium, it is a bit squashed by gravity in the vertical direction, while when it hangs from the thread, it is a bit stretched. Thus, in reality, when the table is removed and the thread is added, the sphere stretches and its center of mass lowers by a finite quantity, and we should take this effect explicitly into account if we wish to discuss the cycle correctly. In the limit $\di T\to 0$ this dropping of the ball remains finite and becomes greater than the raise due to thermal expansion (which tends to zero). Thus, when $\di T$ is small enough, the net result is that after the cooling the center of mass is lower than at the beginning of the cycle, and we cannot easily conclude as before that the second law is violated. A further difficulty is that sudden removal of the table is an irreversible process; if we wish to keep the cycle reversible, the table must be removed slowly while the ball deforms and the force sustaining the ball gradually transfers from being provided by the table to being provided by the thread.

Naively, it could appear that these squashing and stretching effects can be neglected compared to thermal expansion, or at least can be considered independent from it. The cycle argument above shows that this is not the case, otherwise we would violate the second law. Another warning that such effects should not be neglected comes from solid state physics. An alternative way of saying that a ball is squashed and the other is stretched is to say that one is under compression and the other under tension. It is well known that materials can change their heat capacity under compression.\cite{ashcroft} It is also well known that a purely harmonic crystal, i.e. an ideal material made of masses with Hooke's law springs holding it together, would have no thermal expansion at all and would not have a change in heat capacity under compression. Thus, both the change in heat capacity under compression and the thermal expansion depend on the same quantity, namely the anharmonicity of the interatomic potential, and are therefore not independent. The conclusion is that we should take explicitly into account the change in heat capacity under compression.

A re-analysis of the cycle that takes into account all the non-negligible details would be cumbersome, and we do not do it here. If we are interested in calculating a corrected version of the formula \eqref{conventionalresult} it is easier to simply admit that we cannot assume that the internal energy $U$ and the radius $R$ of the sphere depend only on $T$, but must depend also on $g$. When this is done, the analysis below shows that the change in heat capacity due to the presence of the gravitational field can be calculated exactly as a function of the coefficient of linear expansion, see formula \eqref{result}. This calculation also shows that assuming that only $R$ depends on both $g$ and $T$, with $U$ still depending only on $T$, is not enough to derive a correct result and still leads to a violation of the second law. To derive a correct result, it is crucial to admit that $U$ depends on both $g$ and $T$.

\section{Re-examination of the problem}
We now re-discuss the problem to find the correct answer. In order to treat in a common way the two balls, we rephrase the problem as follows. We consider a single ball glued to the table. The case $B$ with the ball hanging from the thread can be recovered by applying a gravitational field pointing upward, and the glue will prevent the ball from flying away. We will neglect any work done on the surrounding air by expansion: we can imagine that everything takes place in vacuum.\cite{fn1}

In the conventional solution the shape of the balls is implicitly assumed to remain always spherical. Instead, our solutions below do not depend on such assumption, and take into account the possibility that the shape of the ball is distorted. We will use $Y$ to denote the height of the center of mass of the ball, which is a well-defined concept for any shape. For the particular case in which the shape of the ball and the symmetry of the density distribution inside it are exactly spherical, $Y$ coincides with the radius of the sphere $R$, but in the more general case it will not have this meaning.

\subsection{Statistical mechanics derivation}
In this section we use statistical mechanics to find out how the heat capacity changes when a gravitational field is applied (with respect to the zero-gravity case). Consider the ball at temperature $T$ glued to the ground.  Let $\hat{H}$ be its Hamiltonian, containing all the interactions between the molecules of the ball, the interaction that keeps glued its lower side to the ground, and gravity. We put a hat on the symbol $H$ to remind us that we are working in the framework of quantum mechanics and therefore the Hamiltonian is an operator, but all the following steps would be valid after a slight change in notation also for a purely classical system. Although the formal steps would be identical, we prefer to work in the framework of quantum mechanics because strictly speaking solids are intrinsically quantum objects that cannot be described classically. For example, in a typical metal the electrons are delocalized on a scale much greater than the distance between two consecutive atoms, a quantum effect that becomes relevant in their interaction both among themselves and with the nuclei and that cannot be neglected if we want to explain macroscopic properties such as electric conductivity. Moreover, if we consider matter as made by nuclei and electrons, classically we cannot even explain why electrons do not spiral down on the nuclei of atoms, blowing up the entire thing.

The equilibrium state of the ball at temperature $T$ depends in general on the Hamiltonian. Changing the gravitational field changes the Hamiltonian, and therefore it changes also the equilibrium state and its properties, such as the equilibrium height of the center of mass (the stronger the gravitational field, the more the ball will be compressed in the vertical direction).

Let $\hat{H}_0$ be the Hamiltonian in the absence of a gravitational field. We can write the Hamiltonian in the presence of a gravitational field as
\begin{equation}
\hat{H}(g)=\hat{H}_0+m\,g\,\hat{Y}\;, \label{Hamiltonian}
\end{equation}
where $m$ is the mass of the ball, $\hat{Y}$ is the operator center-of-mass height, and $g$ is the gravitational acceleration. As noted above, we do not require the ball to remain spherical: the center of mass is a well-defined concept for any shape. Since $\hat{H}_0$ contains all the interactions between particles and the gluing to the ground, it is a very complicate Hamiltonian, but it will turn out that its details are not important.

Most of the following manipulations do not depend on particular form of the Hamiltonian \eqref{Hamiltonian}. In fact, we won't need the explicit form of $\hat{H}(g)$ until Eq. \eqref{explicitH}. Therefore we start by considering a generic Hamiltonian depending on an external parameter $g$, $\hat{H}=\hat{H}(g)$.

\subparagraph{Formula for the Heat Capacity.}
 Define the inverse temperature to be
	\begin{equation}
	\beta=\frac{1}{k_B\,T}\;.
	\end{equation}
We consider the canonical\cite{huang} ensemble, i.e. we assign to each energy eigenstate $|\psi_i\rangle$ with energy $E_{i}$ a probability of occurring
	\begin{equation}
	P_i =\frac{e^{-\beta\,E_i}}{Z}\;,
	\end{equation}
where the normalization factor $Z$ is called partition function. In quantum statistical mechanics it is convenient to characterize the state of the system with the so-called density matrix $\hat{\rho}$. In this formalism, the average of an observable $O$ (i.e. the mean value of the measurable quantity $O$ that we would obtain by measuring many copies of the system, all in the same quantum state $\hat{\rho}$) is given by
	\begin{equation} \label{average}
	\left\langle\hat{O}\right\rangle=\mathrm{Tr}\left(\hat{O}\,\hat{\rho}\right)\;,
	\end{equation}
where $\mathrm{Tr}$ indicates the trace, and $\hat{O}$ is the quantum operator associated with the observable. A system described by the canonical ensemble is associated with the following density matrix:
	\begin{equation}\label{rhocan}
	\hat{\rho}=\frac{e^{-\beta\,\hat{H}}}{Z}\;.
	\end{equation}
The partition function, which is a function of $g$ and $T$, is given by
	\begin{equation}\label{Z}
	Z(g,T)=\mathrm{Tr}\;\left( e^{-\beta\,\hat{H}(g)}\; \right).
	\end{equation}
Defining the following quantity
	\begin{equation}\label{F}
	F(g,T)=-\frac{\ln Z}{\beta}
	\end{equation}
and putting together Eqs. \eqref{average}, \eqref{rhocan}, \eqref{Z} and \eqref{F}, we find that the average value of the Hamiltonian can be expressed as
	\begin{equation}\label{E}
	h(g,T)= \left\langle\hat{H}\right\rangle= \frac{1}{Z}\,\mathrm{Tr}\left(\hat{H}\,e^{-\beta\,\hat{H}}\right)\; = -\frac{\partial\ln Z}{\partial\beta}=F+\beta\frac{\partial F}{\partial\beta}=F-T\frac{\partial F}{\partial T}\;.
	\end{equation}
Normally we would call this quantity (the mean value of the Hamiltonian) the energy of the system, and $F$ would be the free energy. However, we must resist this temptation in order to keep our notation consistent throughout the paper. The reason is that $\hat{H}$ includes the gravitational potential energy, which was not included in what above we called $U$, the internal energy of the system. In this statistical mechanics derivation, $U$ corresponds to the mean value of $\hat{H}_0$ over the ensemble given by Eq. \eqref{rhocan} (where in this latter equation we need to put the full Hamiltonian $\hat{H}$). The quantity $h$ is the mean total energy of the system, the sum of the mean internal energy plus the mean gravitational energy, and it will turn out to correspond to the enthalpy in the purely thermodynamical derivation below. The calculations in this section are however self-consistent and can be understood without referring to the other derivation.

We are interested in what happens when heat is supplied to the system during a process in which the gravitational field $g$ is kept constant. In particular, we are interested in the heat capacity, which is defined as the quantity of heat needed to raise the temperature by one unit.\cite{landau} This clearly depends on the conditions under which the heating takes place: in our case, these conditions consist in that $g$ is constant. Since the table is not moved and the experiment takes place in vacuum, the only external force that can perform work on the sphere is gravity. However, we have already included the gravitational potential energy in the Hamiltonian, so that if $g$ does not change, the heat supplied to the sphere entirely goes to increase the average value of $\hat{H}$ (i.e. the mean total energy $h$).  Therefore the heat capacity is given by
	\begin{equation}
	C_g(g,T)=\frac{\partial h}{\partial T}=-T\frac{\partial^2F}{\partial T^2}\;. \label{eqC1}
	\end{equation}
\subparagraph{Depence of $C$ on the parameter $g$.}
We now consider how the equilibrium state changes when $g$ is varied at fixed temperature. We have
	\begin{equation}
	\frac{\partial F}{\partial g}=-\frac{1}{\beta\,Z}\,\frac{\partial Z}{\partial g}=\frac{1}{Z}\mathrm{Tr}\left(\frac{\partial\hat{H}}{\partial g}\,e^{-\beta\,\hat{H}}\right)=\left\langle\frac{\partial\hat{H}}{\partial g}\right\rangle\;.
	\end{equation}
Using Eqs. \eqref{E} and \eqref{eqC1}, the changes in the total energy and heat capacity when the parameter $g$ is varied can be written as
\begin{eqnarray}
\frac{\partial h(g,T)}{\partial g} &=& \left(1-T\frac{\partial}{\partial T}\right)\frac{\partial F}{\partial g}=\left(1-T\frac{\partial}{\partial T}\right)\left\langle\frac{\partial\hat{H}}{\partial g}\right\rangle\label{dE/dg}\\
\frac{\partial C(g,T)}{\partial g} &=& -T\,\frac{\partial^2}{\partial T^2}\frac{\partial F}{\partial g}=-T\,\frac{\partial^2}{\partial T^2}\left\langle\frac{\partial\hat{H}}{\partial g}\right\rangle\;.\label{dC/dg}
\end{eqnarray}
Using the explicit form of the Hamiltonian, given by Eq \eqref{Hamiltonian}, we have
	\begin{equation} \left\langle\frac{\partial\hat{H}}{\partial g}\right\rangle = m \left\langle\hat{Y}\right\rangle\;. \label{explicitH} \end{equation}
Define $Y$ to be the average height of the center of mass:
\begin{equation}
Y(g,T)=\left\langle\hat{Y}\right\rangle\;.
\end{equation}
Then \eqref{dE/dg} and \eqref{dC/dg} become
\begin{eqnarray}
\frac{\partial h}{\partial g} &=& mY-mT\frac{\partial Y}{\partial T}\label{dE/dT2}\\
\frac{\partial C_g}{\partial g} &=& -m\,T\,\frac{\partial^2Y}{\partial T^2}\;.
\end{eqnarray}
Defining the linear expansion coefficient along the vertical direction (in general temperature dependent) as
\begin{equation}
\alpha(g,T)=\frac{1}{Y}\,\frac{\partial Y}{\partial T}\;,
\end{equation}
we arrive at
\begin{equation}
\boxed { \frac{\partial C_g(g,T)}{\partial g}=-m\,T Y\left(\alpha^2+\frac{\partial\alpha}{\partial T}\right)\;.} \label{result}
\end{equation}
This is the result we were looking for. It states that, in the presence of a weak gravitational field $g$, the amount of heat necessary to raise the temperature by $\di T$ changes, with respect to the case with no gravity, by an amount equal to the right hand side of Eq. \eqref{result} times $g\,\di T$. For most materials,\cite{NixMacNair1,NixMacNair2,wiki} $\pa \alpha / \pa T$ is positive, hence the quantity on the right hand side of Eq. \eqref{result} is negative. Therefore, we need less heat to raise the temperature by the same amount $\di T$ if a downward gravitational field is present compared to the zero-gravity case. The final temperature of the ball on the table will be higher. This result goes in the other direction with respect to the conventional solution.

Note that even though in our problem we allow for non spherical deformations and isotropy is not assumed, the right hand side of Eq. \eqref{result} should be calculated for $g=0$, hence the value of $\alpha$ to be used here is the one in absence of gravity. If the balls are made of a material that is isotropic when no gravity is applied, then the isotropic value of $\alpha$ should be used in the formula. We can also replace $Y$ with $R$ if the balls are spherical in the absence of gravity. The result \eqref{result} is valid up to first order in $g$, under the assumption that this can be treated as a small parameter. Such an assumption could (and should) be verified experimentally by measuring the coefficient $\alpha(g,T)$ for various values of $g$. Finally, notice also that, if $\alpha$ does not depend on $T$, the new result contains only $\alpha^2$ and does not depend on the sign of $\alpha$, while the conventional result does.

\subsection{Purely Thermodynamical Derivation}
A classical thermodynamical $PVT$ system, where the three quantities are pressure, volume and temperature, is a system admitting an equation of state relating these three quantities, and allowing to find any of them as a function of the other two (e.g. an ideal or a  van der Waals gas). We can make an analogy between such a system and our situation. In our case, we expect the equilibrium height $Y$ of the center of mass of the ball to be determined if we know $g$ and $T$. If we increase $T$ (keeping $g$ constant), $Y$ will presumably increase, due to thermal expansion. If we imagine gradually increasing the gravity $g$ (keeping $T$ constant), $Y$ will probably decrease. We can argue that there must be an equation of state of the form $Y = Y(g,T)$ for our system.

We can derive our result \eqref{result} assuming that such an equation of state exists, and that our system obeys the first and second law of thermodynamics, with no further assumptions. The first law is stated as follows:
\begin{equation}
\delta Q = \di U + m\,g\,\di Y\;, \label{firstprinciple}
\end{equation}
where $U$ is the internal energy of the system. The first law states that the heat given to the ball goes partly into internal energy and partly into gravitational energy. The gravitational energy of the system is treated as an external energy and is not included in $U$. This is because it is possible in principle to convert all the gravitational energy into work, by letting the ball fall. If you have any doubt that Eq. \eqref{firstprinciple} correctly accounts for transformations in which $g$ is slowly changed, you can imagine an analogous problem where instead of gravity, each atom of the ball is connected to the ground through a spring whose force is independent of the distance (these springs do not obey Hooke's law). The ball pulls up the strings (in the case $g$ points downward) with a force $mg$. Clearly, from the point of view of the ball there is no difference between this situation and the situation where a gravitational field is present. If you now imagine the ball as a gas pushing up and down these strings, the formula \eqref{firstprinciple} should be clear. You can for example imagine adiabatic transformations ($\delta Q=0$).

The ball could also be able to store energy as internal stresses. This energy would be due to the interaction potential between the molecules, and we choose\cite{fn3} to include such contributions inside the internal energy $U$.

A statement equivalent to the second law is that there exists a state function $S$ called entropy such that
\begin{equation} \di S = \frac{\di U + m\,g\,\di Y}{T}. \label{diS} \end{equation}
In other words, $\di S$ defined in this way is an exact differential.

From the formulas above, we can see that our system is analogous to a classical $PVT$ system where $P$ has been replaced by $mg$ and $V$ has been replaced by $Y$.
In our problem, we want to heat the system keeping $g$ constant. We can express the heat capacity for such a process as
\begin{equation} C_g = \left.\frac{\delta Q}{\di T}\right|_g \label{Cp} \end{equation}
where the subscript $g$ means that we are considering a reversible transformation that keeps $g$ constant. For a $PVT$ system, the quantity \eqref{Cp} would be analogous to the heat capacity at  constant pressure.

The ingredients have all been given, we now need to calculate the result. Consider the quantity
\begin{equation} h = U + mgY\,,\end{equation} which is the sum of internal plus gravitational energy of the ball, and is the analog of enthalpy for a classical $PVT$ system (for which the enthalpy is defined as $h = U + PV$). There is no need to be familiar with properties of this quantity to understand the following. In general, $U$ and $h$ will be functions of two independent quantities among $Y,g,T$. For example, if we consider $g$ and $T$ to be independent variables, $U=U(g,T)$ and $h=h(g,T)$.
The quantity we want, $C_g$, will be in general a function of $g$ and $T$ and can be obtained as a derivative of $h$:
\begin{equation}C_g(g,T) = \frac{\pa h(g,T)}{\pa T} \label{defC} \end{equation}
This can be seen by noting that for a transformation with $\di g = 0$,
\begin{equation} \di h=\di U + m\,Y\,\di g  + m\,g\,\di Y \label{diE} \end{equation}
coincides with $\delta Q$, and then using Eq. \eqref{Cp}.
We now want to calculate $\pa C_g(g,T) / \pa g$, which tells us how the heat capacity changes when a gravitational field is applied with respect to the zero-gravity situation. The proof relies on the fact that $\di S$ in Eq. \eqref{diS} is an exact differential. First note that (combine Eq. \eqref{diS} and Eq. \eqref{diE}):
\begin{equation} \di S = \frac{1}{T}\left(\di h - m\,Y\,\di g\right) \label{diS1} \end{equation}
Expanding $\di h(g,T)$ and using Eq. \eqref{defC} we can rewrite $\di S$ as
\begin{equation} \di S = \frac{1}{T}\left(C_g \di T + \frac{\pa h(g,T)}{\pa g}\di g - mY \di g\right) \label{diS2} \end{equation}
From this last equation we have
\begin{equation} \frac{\pa S(g,T)}{\pa g} = \frac{1}{T} \frac{\pa h(g,T)}{\pa g} -\frac{mY}{T} \qquad \text{and} \qquad \frac{\pa S(g,T)}{\pa T} = \frac{C_g}{T}   \end{equation}
Now calculate both the cross partial derivatives:
\begin{eqnarray}
\frac{\pa}{\pa g}\frac{\pa S(g,T)}{\pa T} &=& \frac{1}{T} \frac{\pa C_g(g,T)}{\pa g} \label{Cross1}\\
\frac{\pa}{\pa T}\frac{\pa S(g,T)}{\pa g} &=& -\frac{1}{T^2}\left( \frac{\pa h(g,T)}{\pa g} - m Y  \right) + \frac{1}{T}\frac{\pa^2 h(g,T)}{\pa T \pa g} - \frac{m}{T} \frac{\pa Y(g,T)}{\pa T}\;. \label{Cross2}
\end{eqnarray}
The second law of thermodynamics (i.e. $\di S$ is an exact differential) is implemented imposing that the cross derivatives commute:
\begin{equation} \frac{\pa^2 S(g,T)}{ \pa T \pa g} =  \frac{\pa^2 S(g,T)}{ \pa g \pa T}.  \label{crossS} \end{equation}
Equating eqs. \eqref{Cross1} and \eqref{Cross2} and noting that
\begin{equation} \frac{\pa^2 h(g,T)}{\pa T \pa g} =  \frac{\pa^2 h(g,T)}{\pa g \pa T} = \frac{\pa C_g(g,T)}{\pa g}, \label{crossC}  \end{equation}
we obtain
\begin{equation} \frac{\pa h(g,T)}{\pa g} = - m \frac{\pa Y(g,T)}{\pa T} T + m Y. \label{almostdone} \end{equation}
This is equivalent to Eq. \eqref{dE/dT2}.
Now as before take the derivative with respect to $T$ of this last equation and use the definition of $\alpha = (1/Y) (\pa Y(g,T) / \pa T)$ and Eq. \eqref{crossC} to recover our result \eqref{result}:
\begin{equation} \boxed{ \frac{\pa C_g(g,T)}{\pa g}  = - m T Y \left(\alpha^2 + \frac{\pa \alpha(g,T)}{\pa T} \right). } \label{result2} \end{equation}

Note that if we made the (incorrect) assumption that $U(g,T)$ depends only on $T$ and not on $g$, which might have been considered reasonable (a similar statement is true for an ideal gas, for example), we would have obtained the conventional result. This can be seen if we note that $C_g$ can be obtained dividing Eq. \eqref{firstprinciple} by $\di T$ for a transformation at constant $g$:
\begin{equation} C_g = \left.\frac{\delta Q}{\di T}\right|_{g}  = \frac{\pa U(g,T)}{\pa T} + mg \frac{\pa Y(g,T)}{\pa T} \label{wrongderivation} \end{equation}
If we assume that $U(g,T)$ does not depend on $g$, then ${\pa U(g,T)}/{\pa T}$ can be calculated in the case $g=0$, and it would also coincide with the heat capacity in absence of gravity, ${\pa U(g,T)}/{\pa T} = C_g(g=0,T)=C_0(T)$. Plugging into Eq. \eqref{wrongderivation} the definition of $\alpha$, we would obtain $C_g(g,T)=C_0(T)+ mg\alpha Y$ which is the same as the incorrect result \eqref{conventionalresult} upon replacing the center-of-mass height $Y$ with the radius of the sphere $R$. Thus, the hidden incorrect assumption made using the conventional solution, as claimed in the discussion of the violation of the second law, is that $U$ depends only on $T$.

\section{Conclusions and final remarks}
In this paper we have re-examined a well-known problem, showing that the conventional solution leads to a paradoxical situation which violates the second law of thermodynamics, and is therefore incorrect. We have proposed a new result, worked out through two different methods. Our solution provides an unusual application of thermodynamics which is reminiscent of the thermodynamics of a rubber band.\cite{feynman,rubberband1}

This problem has two main lessons to teach us. First, it makes us reflect on the definition of temperature. We associate temperature to the average kinetic energy of a molecule (by analogy with an ideal gas or recalling the equipartition theorem), and this may lead to the belief that the internal energy $U$ of the ball does not depend on gravity. This leads to the conventional incorrect solution. However, $U$ does not only include kinetic but also interparticle potential energy contributions. In the context of classical thermodynamics, temperature is defined as the state function such that $\delta Q/T$ is the exact differential of another state function $S$. Only in certain special cases this is related in a simple way to the mean internal energy per particle (which includes kinetic \textit{and} interparticle potential energy contributions). We have seen how our problem is not one of these cases, and the correct definition can be counterintuitive, since the existence of the state function $S$ together with the values of the linear expansion coefficient found experimentally imply that $U$ must depend on $g$ as well as on the temperature.\cite{fn2}

The solution using statistical mechanics provides another way of understanding this. Gravity contributes to the average total energy $h$ in two ways: it adds the average value of the gravitational potential to it, and it changes the equilibrium state since it changes the Hamiltonian. This change in the equilibrium state essentially means that gravity modifies also the internal energy, i.e. $U$ depends on $g$.
A physical way of understanding this goes as follows. At zero temperature and without gravity, the ball is in the minimum energy configuration, i.e. the distance between two adjacent molecules minimizes their interaction potential. As we increase temperature, the molecules will start to oscillate, and since the ball expands, the average distance between two adjacent molecules will be greater than the minimum of the potential, resulting in a greater internal energy. Now we switch gravity on. Consider the ball $A$ on the ground: gravity acts against its expansion, so it decreases the average distance between two adjacent molecules, making it closer to the minimum of the potential. The contrary happens for the ball $B$ on a thread, and we can now understand how $U$ can decrease as $g$ increases in a neighbourhood of $g=0$. This effect goes in the opposite direction with respect to the first (the addition of the gravitational potential), and is completely neglected by the standard solution, that accounts correctly only for the first. However, from the correct solution we know that the two effects are related, and actually the second always wins over the first, so it can never be neglected.

Finally, this problem makes us reflect on the generality of thermodynamics. The idealised problem can be solved without resorting to any approximation, and the solution does not depend on the internal details of the system: nor the spherical shape, nor the atomic-level structure of the ball, nor how it is glued to the ground are relevant. Neither if the world is classical or quantum matters: the first solution uses quantum statistical mechanics, but replacing the traces with phase-space integrals it can be rewritten with classical statistical mechanics and the results will be identical. This fact is more explicit in the second solution, that does not assume nor classical nor quantum mechanics (except for the fact that macroscopic observables do have well-defined classical values): everything is a simple consequence of macroscopic energy conservation and the second law of thermodynamics.

\begin{acknowledgments}
We are greatly indebted to Steve Simon for helpful discussions and for first doubting the correctness of the conventional solution, and to Stefano Bolzonella and Neil Robinson for helpful comments. GdP thanks Marco Polini for useful comments. MCS acknowledges support from a Clarendon Fund Scholarship granted by the Oxford University Press.
\end{acknowledgments}


\begin{thebibliography}{99}
\bibitem{IPhOsite} See the Official International Physics Olympiads website, \url{http://ipho.phy.ntnu.edu.tw}, and in particular the problems of the 1st IPhO at \url{http://ipho.phy.ntnu.edu.tw/problems-and-solutions/1967/1st_IPhO_1967.pdf}

\bibitem{IPhO1} Rudolf Kunfalvi,
\textit{Collection of Competition Tasks from the I through XV International Physics Olympiads 1967 - 1984} (Roland Eotvos Physical Society and UNESCO, Budapest 1985)

\bibitem{IPhO2} Chaleo Manilerd,
\textit{International Physics Olympiads - Problems and Solutions from 1967 - 1995}, (Rangsit University Press, 1996)

\bibitem{200puzzling} P. Gn\"adig, G. Honyek and K. F. Riley,
\textit{200 Puzzling Physics Problems}, (Cambridge University Press, 2001)

\bibitem{madaboutphysics} C. P. Jargodzki and F. Potter,
\textit{Mad about physics: braintwisters, paradoxes, and curiosities}, (John Wiley \& Sons, New York, 2001)

\bibitem{AIEEE} D. B. Singh,
	\textit{AIEEE Physics Arihant: 2006 Edition}, pg. 261, Problem 38. (Arihant Prakashan, 2006)

\bibitem{dukechallenges} Duke Physics Challenges, \url{http://www.phy.duke.edu/~hsg/physics-challenges/challenges.html}

\bibitem{caltech} Caltech Physics League, 2011-2012 Fall Term Game Show, \url{http://www.cmp.caltech.edu/refael/league/2011-2012_fall/Caltech\%20Physics\%20League\%202011-2012\%20Fall\%20Term\%20Game\%20Show.pdf}

\bibitem{princeton} Princeton University Physics Competition, Sample Problems, \url{http://physics.princeton.edu/pupc/files/sampleProblemsOnsite.pdf}

\bibitem{feynman}
See for example R. P. Feynman, R. B. Leighton and M. Sands, \textit{The Feynman Lectures on Physics, The New Millennium Edition, Volume I}, (Basic Books, New York, 2010), \url{http://www.feynmanlectures.caltech.edu}

\bibitem{ashcroft} N. W. Ashcroft and N. D. Mermin
	\textit{Solid State Physics}, (Brooks/Cole, 1976)

\bibitem{landau} L.D. Landau, E.M. Lifshitz 
	\textit{Course of Theoretical Physics, Vol. 5: Statistical Physics part 1, 3rd ed.} (Pergamon Press, 1980)

\bibitem{huang}
See for example K. Huang, \textit{Statistical Mechanics, Second Edition}, (John Wiley \& Sons, 1987)

\bibitem{NixMacNair1} F.~C. Nix and D. MacNair,
	\textit{The Thermal Expansion of Pure Metals: Copper, Gold, Aluminum, Nickel, and Iron} (Physical Review, Vol. 60, pp. 597-605, Oct 1941)

\bibitem{NixMacNair2} F.~C. Nix and D. MacNair,
	\textit{The Thermal Expansion of Pure Metals. II: Molybdenum, Palladium, Silver, Tantalum, Tungsten, Platinum, and Lead} (Physical Review, Vol. 61, pp. 74-78, Jan 1942)

\bibitem{wiki}
See for example \textit{Linear thermal expansion coefficient for some steel grades}, Wikipedia,  \url{http://en.wikipedia.org/wiki/Thermal_expansion#Thermal_expansion_coefficients_for_various_materials}

\bibitem{rubberband1} See for example David Roundy and Michael Rogers,
	\textit{Exploring the thermodynamics of a rubber band} (American Journal of Physics, Vol. 81, Issue 1, pp. 20-23, 2013), and references therein

\bibitem{fn1} It is worthwhile to remark at this point that while we find intuitive to think to the problem using a single ball glued to the table, there are several alternative equivalent ways of thinking the problem setup, all leading to the same set of equations. For example, the reader might prefer to think of flipping the direction of the vertical axis instead of flipping the sign of the gravitational field. With this choice, we would treat the case of the ball $A$ using a coordinate system with the vertical axis pointing upwards and the case of the ball $B$ using a vertical axis pointing downwards. The difference between the two cases can be also thought of as a boundary condition: in the thread case, the top of the ball is the fixed boundary point, while in the table case, the bottom of the ball is the fixed point.

\bibitem{fn2}  Note that for an ideal gas, the value of the thermal coefficient is $\alpha=1/T$ and this does not lead to an internal energy dependent on pressure, which in our case is analogous to $g$. The values of $\alpha$ found experimentally for a metal ball have a temperature dependency significantly different from that of an ideal gas.

\bibitem{fn3} The reader may at this point ask whether it is legitimate to include such terms in $U$, since it could be possible to extract work from them. One can convince her/himself that this is indeed legitimate in the following way. Imagine that the sphere is now wrapped inside an elastic balloon, whose energy stored as surface tension is a function of the surface area of the ball. It is possible to remove the balloon anytime and extract its energy. Let the energy stored as a function of the area be $\Sigma(A)$, where $A$ is the area. At thermal equilibrium, when the ball wrapped inside the balloon has been let relax for a long time, $A$ will be a function of any two of the three independent variables $g$,$T$,$Y$; in the following think $A=A(g,T)$. 

The surface area of the ball, and therefore the energy stored in the balloon, can vary as transformations are performed on the system. Thus, one might want to change the first law to $\delta Q = \di U + mg \di Y\ + \Sigma'(A) \di A$, where $\Sigma'$ denotes the derivative of $\Sigma$ with respect to $A$, and $\di A$ is the change in surface area in the transformation. Accordingly, one should change the definition of the analog of enthalpy to $h=U + mgY + \Sigma(A)$. It is easy to see that Eqs. \eqref{defC}, \eqref{diS1} and \eqref{diS2} would be identical after these redefinitions [use that $\di A$ can be expressed as $\di A = (\pa A / \pa T) \, \di T + (\pa A / \pa g)\, \di g$], and hence the result \eqref{result2} would also look identical. It is understood here that the equation of state $Y(g,T)$, and hence the expansion coefficient, takes into account the balloon and is therefore different from the case without the balloon. The result \eqref{result2} should be used after we have experimentally determined the equation of state, and hence the value of $\alpha$. Eventual stresses inside the sphere can be thought as analogous to the balloon, and it makes no difference to the result \eqref{result2} whether we count them as part of $U$ or not.

The reason why we can consider stresses as part of $U$ is perhaps easier to see in the statistical mechanics derivation. Here, the interactions responsible for the stresses are included inside $\hat{H}_0$, whose average value gives $U$. If we wrap the ball inside the balloon, we would change $\hat{H}_0$, which would modify the equation of state, but not the result \eqref{result}. The crucial point here is that, with or without the balloon, $\hat{H}_0$ does not depend on $g$. This is true because the part of $\hat{H}_0$ that describes stresses inside the ball (or inside the balloon) depends not on $g$, but on the variables defining the microscopic state of the ball (or of the balloon) such as the position of all its atoms. This implies that the partial derivatives of $\hat{H}_0$ with respect to $g$ is zero and allows the derivation to proceed as described in the text.

\end{thebibliography}
\end{document}